*Chapter 9*

# Human-centred home network security

*Derek McAuley[1], Jiahong Chen[1], Tom Lodge[1], Richard Mortier[2], Stanislaw Piasecki[1], Diana Andreea Popescu[2], and Lachlan Urquhart[3]*

This chapter draws from across the foregoing chapters discussing many core HDI approaches and disciplinary perspectives to consider the specific application of HDI in home network security. While much work has considered the challenges of securing in home IoT devices and their communications, especially for those with limited power or computational capacity, scant attention has been paid by the research community to home network security, and its acceptability and usability, from the viewpoint of ordinary citizens.

It will be clear that we need a radical transformation in our approach to designing domestic networking infrastructure to guard against widespread cyber-attacks that threaten to counter the benefits of the IoT. Our aim has to be to *defend against enemies inside the walls*, to protect critical functionality in the home against rogue devices and prevent the proliferation of disruptive wide-scale IoT DDOS attacks that are already occurring [1].

## 9.1 Introduction

The IoT represents a convergence of ubiquitous computing and communication technologies, with emerging uses that actuate in the real world. No longer do ubiquitous computing systems simply sense and respond digitally, now they physically interact with the world, ultimately becoming embodied and autonomous. Hence on top of the legal issues concerned with privacy (see chapter 2), where it is often (contestably) cited that "users don't care", to one of user safety, where users (along with regulators, governments, and other stakeholders) certainly do care. Likewise, industry needs to become aware that this shift also changes the legal basis under which companies need to operate, from one of disparate and often weakly enforced privacy laws, to one of product liability.

The current widely adopted approach in which cloud services underpin IoT devices has already raised major privacy issues. Importantly in an actuated future, untrammelled communications implicating a plethora of heterogeneous online services in their normal operation also brings with it resilience challenges. We must ensure the integrity of actuating systems, which will require greater local autonomy, and hence localised security, alongside increased situated accountability to users.

[1] School of Computer Science, University of Nottingham, UK
[2] Department of Computer Science and Technology, University of Cambridge, UK
[3] Law School, University of Edinburgh, UK

This problem applies in many areas: industrial control, autonomous vehicles, smart cities and buildings, but also includes the intimate and commonly shared context of the home. Importantly, within the foregoing contexts there exist professional IT Network support staff employed to ensure the systems is securely designed, configured and operated, and hence the underlying technology and its management is targeted at professional users with technical background and training. Likewise, such systems often limit device procurement, enforce device registration for network access and mandate updates and patches be applied before allowing a device to communicate on the relevant network.

It is a very different situation in the home, where the majority of the deployments will not benefit from any specific technical knowledge or training of the users, while currently most home installations place any devices plugged in or having gained the WiFi password access immediately onto the same single network, providing no isolation between attached devices and the traffic they generate.

With this in mind we created the DADA project (Defence Against the Dark Artefacts) to investigate how we could effectively provide usable network security for the now nearly ubiquitous local area network in peoples' homes. We have adopted an approach deeply rooted in pragmatism that recognises the 'real world, real time' conditions that attach to the IoT in the home:

- that the cyber security solutions currently being defined for IoT systems will not deal with legacy issues and will never achieve 100% adoption;
- that extant businesses limit the period of time for which they will provide software and security updates (even if they don't go out of business);
- that cyber security is an arms race and threats will continue to emerge in future;
- and that the public will never become network security experts.

> **Assisted Living**
> The scenario of assisted living helps to illustrate the goals of this research. Imagine a domestic future in which people are mixing devices with substantially different reliability requirements. Firstly, there are critical devices that support the delivery of healthcare services where interference in their normal operation could be life threatening. Then there are devices that support building management and domestic security where interference could be extremely disruptive and costly. The integrity of these devices and their communications must be ensured. Add to this connected domestic appliances that support a wide range of mundane activities where interference amounts to little more than an annoyance - people will not die, or even be seriously affected by such interference, merely inconvenienced - even so, users still want to know what is happening in their network when things go wrong.

Isolating and protecting IoT devices requires that we develop and apply edge learning technologies (as covered on chapter 8) to generate and evolve *models that profile device and inter-device behaviour*. These models will, in turn, implicate the capabilities of the switching and routing fabric that tightly constrains network communications in the home.

Key to the network design transformation is a *human-centric view of infrastructure* to ensure that secure systems providing the greatest degree of isolation between devices do not become burdensome to users, but rather are *accountable* and *socially relevant* [2] and thus provide the legibility, agency, and negotiability required [3] to enable people to exercise control over IoT device behaviours and communications in their everyday lives.

Our focus then is on the evolution of the sociotechnical home network context into which domestic IoT is being deployed. This approach builds upon work from both the Databox home server infrastructure [4] and Homework user experience design methodologies [5], while looking at the new challenges of intelligent home cyber-defence.

## 9.2 Networking background

The fundamental network security concepts we adopt in DADA are already widely deployed in enterprise class networks: separation of functionality and isolation o devices by use of VLANs, separate IP subnet addresses spaces and enforcement of "Guest" status in networks so that devices are only able to send packets to the default router, and not even talk to other devices on the same layer 2 network. Likewise, traffic enforcement of both destination addresses is standard issue firewalling technology, while traffic shaping to limit bandwidth use is widely deployed in integrated communications infrastructure that mixes both data and voice traffic.

Within the DADA project the challenges are then the specific requirements of domestic IoT, e.g., understanding the varied communication patterns of clusters of devices - like peer-to-peer or device-to-hub, in addition to use of broadcast discovery mechanisms. These requirements then need to be mapped to specific switching, routing and firewalling mechanisms that can provide segregated sub-networks. Importantly, where shared resources are used (e.g., upstream Internet links, in-home links carrying multiple subnetworks), traffic shaping and rate control will enforce traffic profiles, with specific attention to rate limiting actuation events, which may involve compound network interactions, that could lead physical devices to be particularly susceptible to denial of service by excessive use [6] or operation outside of design envelope [7]. Traffic shaping will also be available on a per device basis, applying shared profiles created by the ubiquitous monitoring and made available through the profiling service, providing both traffic management and attack detection

These enforcement mechanisms also need to be complemented by device profiling, where we seek to learn and share behavioural and communications patterns of IoT devices – such systems need to be lightweight and accurate, and being based on edge-based sampling techniques and modelling, engage in pseudonymous profile.

General characterisation of Internet-connected devices is known to be complex [8,9]. However, domestic IoT devices have much simpler networking properties than general Internet-connected general-purpose computers, and the scale and degree of multiplexing is thus much smaller and amenable to enumeration and computation. While manufacturers could, in principle, supply device behaviour profiles (e.g., via a Hypercat endpoint) to provide a priori information to management systems, devices deployed in these environments are likely to have long lifetimes, and will undergo many firmware/software upgrades resulting in evolving behaviours, as will the natural evolution of the cloud services used to support such devices. Continual monitoring and re-profiling will thus be necessary. Furthermore, specific attention will be given to identifying physical actuation interactions, potentially involving compound network interactions with multiple devices. In addition, homes will often contain only a small number of instances of many devices (e.g., smart TVs, fridges, electricity meters) rendering it necessary to *share* information across a statistically representative number of monitored networks to build an evolving dynamic model of behaviour while being sure to preserve the privacy of individuals and avoid revealing personal information through sharing device behaviours [9].

Building new technical capabilities into home network infrastructure inevitably leads to questions of how do users interact with such systems and how are they made accountable to end-users, in socially relevant ways that enable end-users to exercise control over their network.

### 9.2.1   Network mechanisms

In order to provide the basis for traffic interventions we must first monitor traffic on the local network and provide a means to identify anomalies. Here we present eBPF-IoT, a customized system for securing Internet-of-Things smart homes – in a future smart home this would be running continuously within the home router, or in more complex domestic networks within all active networking elements. Our specific implementation here is based on Linux.

The eBPF-IoT is comprised of two components: i) eBPF-Mon, which computes traffic statistics and extracts machine learning features needed to train machine learning algorithms for IoT device identification and anomaly detection; and ii) eBPF-IoT-MUD - a traffic management module based on eBPF which implements traffic policies derived from Manufacturer Usage Description (MUD) files provided by the IoT manufacturer.

The extended Berkeley Packet Filter (eBPF) provides a very general way to hook code into network data paths within the Linux kernel without modifying the kernel source code. To ensure system stability the eBPF framework ensures that the code loaded is both safe to execute and loop free (and so will terminate). Our eBPF-Mon component then uses these hooks to gather statistics about flows in real-time.

The eBPF program can hook into the eXpress Data Path (XDP) to efficiently implement packet dropping, for example, to block a Denial-of-Service (DoS) attack - in our scenario we are specifically interested in blocking such traffic originating from devices within the home, whether a DoS on another device in the home or as part of a Distributed DoS targeting a service on the Internet. More complex controls are also possible, such as rate limiting flows, or blocking / forwarding packets at a fine grain, for example, specific ports on specific servers. In contrast to general purpose computers the Internet services that an IoT device would connect to in normal operation are fixed and known by the manufacturer, and the MUD specification (RFC8520) defines how these services can be listed, and in an extension, the expected traffic rates to each service when in operation. So, our eBPF-

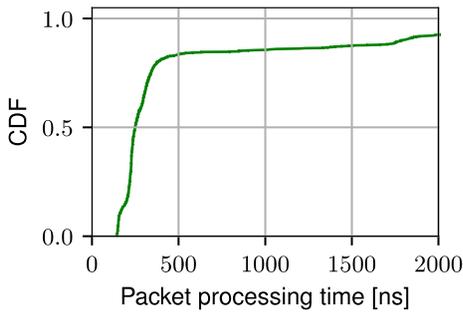

*Figure 9.1 Additional Delay in packet processing time*

IoT-MUD component, uses the MUD profiles of devices to configure the XDP to only permit certain communications from the IoT device (and block all others) and in combination with the eBPF-Mon enforce rate limits on the network flows.

Performance measurements show that we can sustain over 1 Gbps in most scenarios, with only modest added delay to processing the packets in the router.

## 9.3 Shaping User Interaction

To what degree do users *need* to interact with home security systems? As we have seen in previous discussion, monitoring and mitigation of network security threats can be handled transparently by the underlying system; developments in device profiling, pattern recognition and the network control plane can increase the ability of a system to effectively respond to, for example, misconfigured, malicious and poorly designed IoT devices. There is a strong argument for involving the user as little as possible, if at all, in the resolution of these problems, to allow security experts and systems designers to make the security choices for end-users; indeed [10] have argued that the network should be an unremarkable and mundane part of a household; that foregrounding it to users may run the risk of making networks "constantly remarkable and thus problematic". Two factors, however, argue against fully cocooning a user from the concerns of their network security:

1. Home network security pervades and is deeply entangled within the social fabric of a household. Privacy, for example may be equally violated within the home as without. Similarly, the management of access to resources (devices, filesystems, printers and so forth) are subject to routine matters in the home, such as working or socialising.

2. Security management may often be subordinate to the here and now of a household occupant's activities. Where remedial actions taken by a network are disruptive (for example consuming bandwidth, rebooting devices or restricting or even disabling connectivity), automated fixes may be at odds with the immediate concerns of members of the household.

Our challenge, therefore, is to design interactions that support users in the management of the security practices mentioned in (i) whilst ensuring that the home network remains 'unremarkable'. Security-based interactions must defer those out-of-band priorities that arise from day-to-day living. Put another way, we must design with an awareness of the typical philosophy that *"some kind of working is good enough for now"* [10]

Rules and policies are a key feature of network systems, and one approach to folding in social concerns to the management of home network security is to have the system capture and recognise those events (temporal, system, and household) that form the basis of a householder's reasoning about how their system should behave. Temporal events might be 'between 10 and 11 in the morning', system events might be 'using social media, or device uploading 10% more data than normal' whereas household events might be 'when Mum gets home' or even 'Jamie is watching TV'. Ideally, events might be composed to reflect the nuanced needs of household, the used to trigger appropriate system behaviours. Though a beguiling notion, in practice there are a set of (perhaps intractable) issues that limit its use; not least the thorny technical challenges of capturing and codifying higher level constructs (such as ownership, activities such as 'shopping' or 'doing homework') into sane system primitives. Moreover, given that "*different people are allowed to do different things with different devices at different times of the day and in different places*" [11], it is simply too difficult for users to anticipate and express rules to cover even a subset of scenarios; studies have shown that in reality, rules are established retrospectively in response to experience (ibid).

The authors of [12] argue more expansively against relying on fully automated systems security, suggesting that there are inherent limitations to how well automation can succeed in practice *even if the technology behind it is faultless*. They suggest that these limitations can lead to "failures" of automation that are not only technical (i.e., when the automation system simply stops working, for example), but are also failures of meeting the actual needs of the users. Moreover, as the authors of [13] suggest, we "*Only need to make the wrong decision once to discourage future use*".

Given that fully automated approaches to system security are unlikely to provide a complete solution, we continue to require mechanisms to help users interact with their network security; these will include manually responding to issues, restricting or elevating resource access, updating firmware or software, monitoring behaviours and usage. The prevalent support provided today is by graphical user interfaces; these vary in complexity and functionality but will typically provide a degree of *monitoring* and *control*. As the range of security vulnerabilities increases alongside the number of devices in our homes, there may need to be a corresponding increase in the complexity of the interfaces we use to monitor and control them. These interfaces may even come to resemble enterprise systems, though these are often simply "*too complex for home users*" [14]. In response, alternative approaches have been developed and studied. The authors of [14], for example, provide spatial and logical views of a user's home that can be directly manipulated. Others mimic already understood home devices such as alarm or heating panels [15] and we've even seen standalone physical, single purpose devices built to simplify traditionally problematic tasks [16][17].

Outside technical tools for controlling the network, there is some evidence that householders use non-technical approaches to assume control; access control, for example, may be managed by negotiation, though where enforcement is required, more imaginative approaches have been observed, such as locking devices away, or the deliberate creation of 'dead zones' in the house where broadband is unavailable due to an access point being out of range [10]

### 9.3.1 Designing for control

To address the observations and challenges we have discussed, and to walk the line between on one hand keeping the network 'unremarkable' and on the other providing adequate interaction support when required, we have built a mobile and physical interface to facilitate a further examination of a householder interactions with their devices and networks. In the first instance, we have designed mechanisms for users to *control*, rather than *monitor* network security. We have limited the scope of 'control' to a range of simple actions: rebooting devices, restricting access to resources, device monitoring and switching between privileged and unprivileged networks. We are explicitly assuming that *all* householders may wish to perform network control actions on one or more sets of devices, but that social constructs, such as ownership and household hierarchy (rather than, for example, technical competence) can be respected when determining the set of control mechanisms that are available. We have built a physical component for several reasons; first, it is *situated*, meaning that we can explicitly require that a householder is present to perform some tasks, and that the act of performing a task can be made visible to other occupants. Second, it enables us to simplify the interactions and corresponding effort of performing control tasks (so use is predicated upon little to no technical competence). Third, it extends the opportunity for householders to elevate or remove control mechanisms, using non-technical and familiar means (for example, moving out of reach, hiding, locking away, loaning); something that we are especially interested in studying.

We use physical tokens as proxies for home devices (Figure 1.1); an approach reminiscent of the interactive 'tokens and constraints' work first proposed by [18]. The tokens use MiFare[4] 1K RFID tags to communicate with an Android smartphone and a reader. The reader consists of a microprocessor (Adafruit Feather M0 Wifi)[5] that is connected (over $I^2C$) to an (MFRC522[6]) RFID reader, which is capable of reading multiple tags simultaneously. The microprocessor has a wifi chip and communicates with a MQTT broker running on our home access point.

---

[4] https://www.mifare.net/en/
[5] https://www.adafruit.com/product/3010
[6] https://www.nxp.com/docs/en/data-sheet/MFRC522.pdf

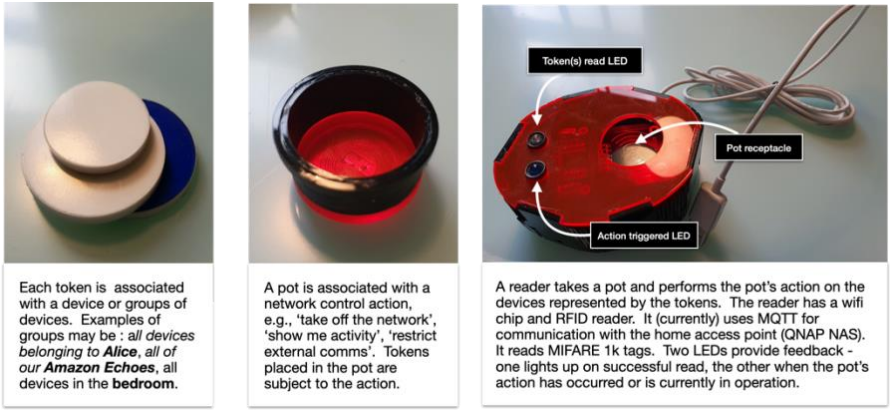

Figure 9.2  Tokens, pot and reader.

Those that are in receipt of tokens will have control of the corresponding device(s) that the token represents. Crucially, tokens can be associated with a single device, or arbitrary device groups that make sense to a particular household (e.g. all of Ellie's devices, all televisions, the thermostats in the top bedroom). Tokens can be replicated, so that users can share control, and they can be modified to add or remove devices as required.

To perform a security-related task, tokens are placed into a 'pot'; a pot is associated with an action, such as 'remove from the network' or 'provide access to printer' or 'log all traffic'. The pot also has an RFID tag but is encoded to distinguish it from tokens. When a pot is placed into a reader, all of the tokens that are in the pot will be subject to the associated action. The reader will light up an LED to show that the action has been triggered. Actions may be *continuous* or *discrete*. That is, in some cases, removing tokens from the pot or reader will undo the action (for example revoke permissions, stop traffic logging), in other cases, the action will remain until superseded, by placing the token in another pot; this allows a user to compose sets of actions whilst keeping hold of the token. Note that, once in receipt of user study data, we may eventually choose to exclusively support only one of these two modalities.

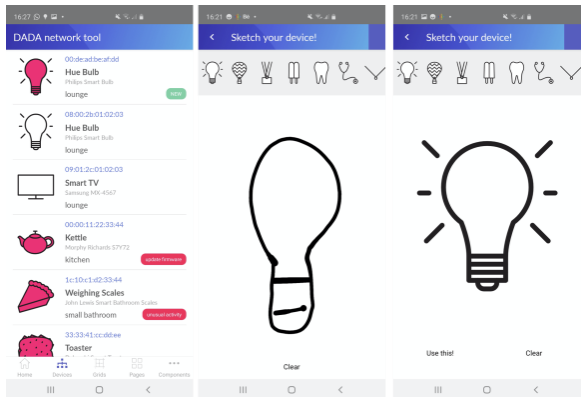

Figure 9.3  App device introduction.

Our mobile app is used to bootstrap and configure our tokens and pots. The app presents a list of devices on the home network, and, using Google's quick draw API[7], provides a simple interface for sketching the device (to help users distinguish between all of their devices) (Fig 1.2).

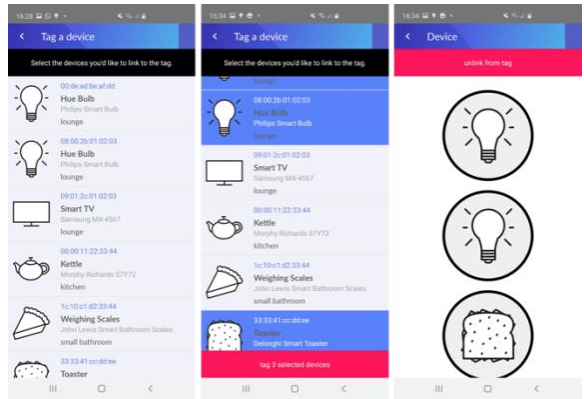

*Figure 9.4  Association of tokens with devices.*

To associate tokens with devices, a user holds a token against the mobile phone. If the token is not already associated with any devices, the user is invited to select a set of devices to associate it with (Fig 1.3). The mac addresses of the devices are then written to the token's tag. If the token already has devices associated with it, the user will be shown the devices, and can remove them and re-associate if required.

The process of configuring pots is similar. To configure a pot, a user holds it against the phone, and they will then be presented with a set of system activities that they can have the pot perform (Fig 1.4). Again, if the pot already has action associated with it, the user can remove them and add alternatives.

---

[7] https://quickdraw.readthedocs.io/en/latest/api.html

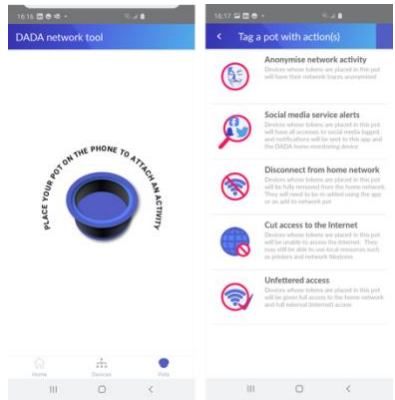

*Figure 9.5 Association of tokens with devices.*

To begin with, we will only support a small set of actions, though this can be expanded in the future. Where it makes sense, the system will allow actions to be composed and associated with a single pot. By designing our system in this way, with programmable pots and configurable tokens, our intention is to enable users foreground and simplify problematic or frequently needed functionality (for example, rebooting a router, dealing with visitors, revoking access to social media) whilst ignoring (or leaving the system to deal with) any features or devices that are or little or no interest. Though the system is principally designed to be reactive, the actions associated with pots can themselves be made contingent upon events if required.

## 9.4 Compliance by Design

One of the challenges facing smart home technologies is the uncertainties within the regulatory environment where multiple forms of normative enablers and constraints may come into play, such as design codes, technical standards and law. At the same time, being a concept for a future domestic IoT security solution, DADA is by definition a smart home technology itself, and will thus be subject to a similar set of challenges as well. To ensure smart technologies – including security solutions, such as DADA – are designed to be truly usable by non-expert users in their everyday life, it is crucial to understand how compliance requirements have affected design, and how design can support compliance.

We take the view that 'secure by design' [19] is not just about bringing the development of smart technologies in line with the security requirements, but also about creating the right conditions for law and technologies to co-evolve in such a way that all involved parties are incentivised to be on board – what one might call the 'legal design' [20]. In this regard, reaching out to different sectors throughout the entire project lifecycle, so as to understand the regulatory landscape as well as the stakeholder expectations, forms an important part of our work at DADA. Through working with the academic, industrial and policymaking communities, we have envisaged, discovered, verified and pre-empted some of the design and compliance challenges facing the industry as well as DADA.

### 9.4.1   *Outlining the regulatory framework with academic research*

In the scholarly literature, while there have been a growing body of discussions in improving privacy and security in the IoT environments, such efforts have largely focused on the technical aspects of the IoT artefacts or the interactions between such artefacts and human users. The regulatory environment, though touched upon as a relevant topic, has been mostly framed as either the motivation for research initiatives or the external constraints for research and development. In developing our work at DADA, we take a bidirectional approach by unpicking not only how the technological paradigms for IoT technologies can shift to improve compliance with the existing regulatory framework, but also how the regulatory environment can evolve to support the human-centred approach we advance. By 'regulatory environment', our work mainly addresses two normative realms: law and standards.

The legal thread of our investigation deals with the data protection legal framework, mainly the GDPR. Considering the sheer volume and potential sensitivity of the data collected and analysed by IoT devices, including DADA as a future IoT cybersecurity solution, the starting point of our inquiry is whether and to what extent data protection law applies to edge computing in smart homes. In our paper 'Who is responsible for data processing in smart homes?' [21], two key legal concepts have been foregrounded to highlight the impact of data protection law on the development and adoption of security- and privacy-enhancing technologies: joint controllership and the household exemption. We note that these two notions are not simply static, immutable rules, but rather an ever-adapting pair that have been shaped up through regulatory guidelines and case-law over the years. The expanding scope of joint controllership – meaning developers and end-users of smart technologies are more likely to be held responsible as joint data controllers – and the shrinking household exemption – meaning end-users are less likely to be exempted from data protection obligations – will collectively create a regulatory environment where the research and uptake of IoT privacy and security technologies may be deterred by the potential compliance burdens. These trends are well-intended with the objective to provide seamless protection to data subjects, but have failed to recognise the nuances of control and power in digital domestic life. We therefore concluded that, '[p]aradoxically this may then result in a lower degree of privacy, as well as security, for smart home inhabitants.' [21]

This message is further elaborated in Chapter 2 of this book, where we further unpack the legal principle of 'accountability' in a smart home environment and also in a more user-centric approach. This is achieved through the construction of the notion of 'domestic data controllers' and the advancement of the differentiated responsibilities in smart homes. These discussions are relevant to our approach with DADA because it underlines how the current data protection legal framework does not fully capture the contemporary reality of digital life that is characterised by the mundanity of ordinary people exercising control over personal data as well as the complex techno-socio-legal dynamics behind the decision-making on IoT privacy and security issues by stakeholders. To support domestic data controllers to demonstrate accountability, we argued, it would be essential to make sense of the interpersonal relations in the digitalised domestic politics, the role of technologies in this domain, and the differentiated nature of power exercised by domestic controllers and vendors. All these have profound implications for policymaking, including how the regulatory environment may be improved with clear and fair guidance.

A second thread of our work focuses on cybersecurity standards. We have written on how these standards are currently showing a similar lack of acknowledgement of the latest socio-technical developments in smart homes [22]. For one thing, major technical standards for IoT security are predominantly developed with the stereotypical cloud-based architecture in mind. The recurring recommendations of those standards on how smart devices should be designed (for example, the resilience to outages requirement) would be often also relevant for edge-computing architectures and actually more efficiently implemented by the latter. While cloud-based standards are needed considering the number of IoT devices relying on this architectural model, standards can also set out directions for best practice and, therefore, should be more aspirational in attempting to change current industry practices, especially in the context of the rapid developments in edge-computing. For another thing, current standards focus mainly on technical and external threats, leaving a significant gap in addressing the internal human threats in smart homes. For example, domestic abusers may exploit IoT devices to monitor, harass, intimidate, control or manipulate other members of the household, but most, if not all, of the current cybersecurity standards have failed to provide guidance on how to anticipate and minimise such risks. Standards could potentially help in removing affordances related to in-home human threats and further research in this field is required. In this regard, the routine activity theory (RAT) can be a helpful conceptual framework for uncovering the values of edge-computing and translating such strengths into actionable strategies to defend against security threats, whether technical or human in nature, and whether from within or outside the home [23]. Again, the shortcomings of the current standards will have policymaking implications, especially the urgent need to deepen our understanding of different types of security risks, and the roles and limitations of technical standards.

Through our research into the status quo of data protection law and cybersecurity standards, we aim to communicate a message to the academic community, one that echoes the underlying idea of DADA: Human-centred IoT security is not just about technological design, but also about normative development. A supportive regulatory environment, created with the right set of legal rules and technical standards, will promote security in smart homes. For academics and policymakers alike, this means further interdisciplinary work is needed to comprehend the human factors in compliance, design and regulation in the area of domestic IoT. The policymaking implications of DADA will be further discussed in Section 8.7.3 below.

## 9.5 Extracting practical insights from industrial involvement

In conjunction, we want to ensure that our work engages with practical challenges of designing for cybersecurity in IoT development. As such, we focus on two strands of our work here, where industrial insights have been key in charting the problem space.

Firstly, we conducted a series of exploratory semi structured interviews with different stakeholder groups from the IoT sector. Whilst industry practice around IoT security was a key focus, we also wanted to understand the wider IoT ecosystem and thus spoke to NGOs and government bodies. This provided us a wider appreciation of the emerging threat landscape, priorities for industry, strategies for anticipating and managing these risks and what direction the industry is headed in securing IoT.

For example, in relation to motivations and solutions. The motivations for building secure IoT systems ranging from fear of harder regulation and reputational harm to emerging threats, such as how to manage vulnerabilities at scale (e.g. a washing machine bug in thousands of homes leading to a critical national infrastructure threat, not just an individual one). In terms of solutions, edge computing, greater need for life cycle security management, value of MUD profiles and use of labelling of IoT risks all emerged as valuable. Such insights can guide academic research, providing interesting problems to be addressed.

Secondly, we are aware that developers have a growing role in addressing socio-technical harms of IoT systems, as seen in the Secure by Design work discussed below. Thus, we wanted to create a tool to help translate and make discussions around IoT cybersecurity more accessible. Acknowledging the value of reflective tools for thinking about law and ethics (e.g. ideation card decks) [24], we wanted to explore the value of play, and developed a 'serious board game' for helping IoT developers consider human dimensions of securing smart homes. Conscious of moving past the framing of purely technical solutions, the goal was to create a game that engages with the complexities of managing home networks, guarding against a hacker who is compromising IoT devices. The game uses play to raise awareness about appropriate strategies to secure the home, nature of risks and challenges domestic users face in using different skillsets to address these dangers. The game has been developed through a process of iterative development with different stakeholder groups from experts in cybersecurity and games, to those working on IoT, HCI and usable privacy and security.[8] As such, our process has been shaped by their insights, where industry, as well as academic expertise, has played a key role in guiding gameplay, aesthetic, and themes covered. Furthermore, with Covid-19, we needed to adapt our planned game development approach. Originally it was to be a physical game but we still managed to use Tabletopia (an online boardgame platform), to iterate and test our prototype, to great success.

## 9.6 Shaping the prospect by informing policymaking

DADA started as a research project at a time when large-scale cybersecurity attacks have raised cross-departmental concerns in the UK government [19], followed by the publication of a code of practice for consumer IoT security [25], and later a set of regulatory proposals [26]. At the time of writing, the Department for Digital, Culture, Media & Sport (DCMS) has conducted a formal consultation on its initial positions (May-June 2019) [26], feeding into its revised proposal open for a second round of comments (July-September 2020) [27]. In a separate but related review, the DCMS also carried out a review on cybersecurity incentives and regulation, with a call for evidence (November-December 2019) [28]. This highlights the importance of regulating IoT cybersecurity as a governmental priority, and also represents an opportunity for DADA to translate research findings into something that can potentially impact the regulatory environment surrounding smart technologies.

---

[8] Many thanks to our project intern Adam Jenkins, PhD researcher in usable security at School of Informatics, University of Edinburgh. He has been key in the game design and iterative testing process.

We have submitted our responses to all three inquiries [29, 30, 31] with a view to raising some of the legal issues we have identified in the journal articles and the practical considerations flagged up in our engagements with industrial stakeholders. Responding to public consultations, however, has the limitation that these consultations always take the form of a set of pre-determined, sometimes strictly structured questions, and occasionally, there is a preferred position, be it explicitly specified in the consultation document or inferred by the way the questions have been framed.

Exposing the 'blind spots' of these inquiries can therefore be somewhat challenging, because going beyond the scope of the questions may result in the answers being disregarded. From the three IoT-related inquiries initiated by the government, there is a clear lack of the emphasis on the human aspects of security in smart homes. The regulatory proposal for the May-June 2019 consultation, for example, views the regulatory challenge being 'vulnerable devices becom[ing] the weakest point in an individual's network' and '[c]ompromised devices at scale [posing] a risk for the wider economy through distributed denial of service (DDOS) attacks such as Mirai Botnet in October 2016.' [32] Accordingly, the first proposed regulatory option – which became the selected model in the July-September 2020 version [33] – focused entirely on the three technical safeguards: 'no default passwords', 'implement a vulnerability disclosure policy', and 'keep software updated'.[9] Our response highlighted the need to take into account the user-friendliness consideration of managing IoT security mapping these design requirements to the existing cybersecurity guidelines, such as the 'make installation and maintenance of devices easy' principle in the government's own code of practice. We submitted that 'most of the ten guidelines additional to the "top three" are already legally required under data protection law and cybersecurity law' and therefore mandating all 13 principles 'would not *per se* create a significant or disproportionate amount of compliance costs to retailers or manufacturers' [29].

As regards the organisational dimension of addressing cybersecurity threats, the July-September 2020 review has touched upon the lack of ability and motivation for businesses to put in place security measures. Many of the consultation questions, however, are based on the assumption that the level of incentives and barriers varies along the spectrum of size of business (e.g. small, medium and large organisations). Yet, through our expert interviews, we found that organisation size is not the only decisive factor, other parameters such as the organisation's position in the supply chain, the market structure of a particular sector and the business model can also significantly affect how motivated a business can be to manage cybersecurity properly. As such, we submitted that, '[a]n effective cybersecurity strategy will need to acknowledge the heterogeneity within and across sectors on those dimensions.' [28]

There is clearly a delicate balance between foregrounding the under-discussed issues and remaining in scope when it comes to responding to governmental consultations. DADA as a research project has endeavoured to communicate the message of a human-centred approach in cybersecurity to policymakers by carefully aligning this with how the challenges and solutions have been framed. For future projects, other forms of engagement, such as a more proactive initiative to contribute to the scoping of the consultation in an earlier stage, may prove more effective if the timing of the project lifecycle allows.

---

[9] These are known as the 'top three' principles in the Code of Practice for Consumer IoT Security. See Department for Digital, Culture, Media & Sport. *Code of Practice for Consumer IoT Security* [online]. 2020. Available from https://www.gov.uk/government/consultations/consultation-on-regulatory-proposals-on-consumer-iot-security [Accessed 26 Sep 2020].

## 9.7 Lessons of DADA on compliance by design

Our work on compliance by design in domestic IoT has not only enabled us to ensure DADA as a research project is privacy-aware, value-driven, ethically-sustainable and legally-compliant, but more importantly to explore approaches to understanding and interacting with the surrounding regulatory spaces that matter to future developments of security- and privacy-enhancing technologies for smart homes. Through different threads of research and outreach activities with the academic, industrial and policymaking communities, some of our experiences may be of value to future research projects:

Firstly, for interdisciplinary projects, there can be a strong role for social scientists to play in co-shaping the technical approach, provided that messages are clearly translated across discipline-specific languages. The initiative to proactively engage with audiences coming from a more technical background is key.

Secondly, with a multi-stakeholder approach, it is essential that the priorities of different groups are clearly identified so as to inform the adjustment of the engagement approaches. Industrial partners, for example, would have starkly different motivations behind improving IoT security from policymakers. Consequently, in order to generate greater impact, it would be helpful to explain how the human-centred approach is in line with their own overall agenda.

Thirdly, managing interdependencies and continuity across different strands of work has proved significantly crucial. For example, the RAT theory employed in our academic work has ended up being the conceptual framework for the design of the serious game. Also, the findings in our interviews with experts from the industry have also laid the empirical groundwork for our policy engagement activities. Effective communications between, and timely reflections on, all fronts of the work have made this much more manageable.

Fourthly, with a large part of our work taking place during the COVID-19 pandemic, adaptability and resilience have turned out especially important to our empirical research. While the plans to engage with industrial stakeholders through interviews and the game were affected by the restrictions, eventually it was possible to move these activities online. It also offered us the opportunity to reflect on how different groups of stakeholders have responded to our new approaches, and what the more effective ways to engage with them would be going forward.

## Acknowledgement

This work was supported by the Engineering and Physical Sciences Research Council [grant numbers EP/M001636/1, EP/N028260/1, EP/M02315X/1, EP/R03351X/1]. This chapter is based on the following original works:

- Piasecki, S., Urquhart, L., and McAuley. D. (2021) Defence Against the

Dark Artefacts: Smart Home Cybercrimes and Cybersecurity Standards. *Computer Law & Security Review: The International Journal of Technology Law and Practice*. (In Press).

- Chen, J., Edwards, L., Urquhart, L., McAuley, D. (2020) Who Is Responsible for Data Processing in Smart Homes? Reconsidering Joint Controllership and the Household Exemption. *International Data Privacy Law*. 2020; 10(4) pp. 279-293.

# References

1. Symantec, https://www.symantec.com/connect/blogs/iot-devices-being-increasingly-used-ddos-attacks
2. Privacy-aware infrastructure for managing personal data, http://dx.doi.org/10.1145/2934872.2959054
3. HDI?
4. Amar, Y., Haddadi, H. and Mortier, R. (2016) "Privacy-aware infrastructure for managing personal data", *Proceedings of SIGCOMM*, pp. 571-572, Florianópolis, Brazil, ACM Press. https://doi.org/10.1145/2934872.2959054
5. Crabtree, A., Rodden, T., Tolmie, P., Mortier, R., Lodge, T., Brundell, P. and Pantidi, N., 2015. House rules: the collaborative nature of policy in domestic networks. Personal and Ubiquitous Computing, 19(1), pp.203-215. https://doi.org/10.1007/s00779-014-0771-6
6. A Survey: DDOS Attack on IoT http://www.ijerd.com/paper/vol10-issue11/Version_3/I10115863.pdf
7. The real story of Stuxnet https://spectrum.ieee.org/telecom/security/the-real-story-of-stuxnet
8. Guide to intrusion detection & prevention systems, http://dx.doi.org/10.6028/NIST.SP.800-94
9. Spying on the Smart Home: Privacy Attacks and Defenses on Encrypted IoT Traffic, https://arxiv.org/abs/1708.05044
10. Crabtree, A., Mortier, R., Rodden, T. and Tolmie, P., 2012, June. Unremarkable networking: the home network as a part of everyday life. In *Proceedings of the Designing Interactive Systems Conference* (pp. 554-563).
11. Crabtree, A., Rodden, T., Tolmie, P., Mortier, R., Lodge, T., Brundell, P. and Pantidi, N., 2015. House rules: the collaborative nature of policy in domestic networks. *Personal and Ubiquitous Computing*, *19*(1), pp.203-215.
12. Edwards, W.K., Poole, E.S. and Stoll, J., 2008, July. Security automation considered harmful?. In *Proceedings of the 2007 Workshop on New Security Paradigms* (pp. 33-42).
13. Bauer, L., Cranor, L.F., Reiter, M.K. and Vaniea, K., 2007, July. Lessons learned from the deployment of a smartphone-based access-control system. In *Proceedings of the 3rd Symposium on Usable Privacy and Security* (pp. 64-75).

14. Yang J., Edwards K., and Haslem D. (2010) "Eden", Proc. of UIST, pp, 109-118, New York, ACM
15. Mortier, R., Rodden, T., Lodge, T., McAuley, D., Rotsos, C., Moore, A.W., Koliousis, A. and Sventek, J., 2012, January. Control and understanding: Owning your home network. In *2012 Fourth International conference on communication systems and networks (COMSNETS 2012)* (pp. 1-10). IEEE.
16. Yang, J. and Edwards, W.K., 2007, September. Icebox: Toward easy-to-use home networking. In *IFIP Conference on Human-Computer Interaction* (pp. 197-210). Springer, Berlin, Heidelberg.
17. Balfanz, D., Durfee, G., Grinter, R.E., Smetters, D.K. and Stewart, P., 2004, August. Network-in-a-Box: How to Set Up a Secure Wireless Network in Under a Minute. In *USENIX Security Symposium* (Vol. 207, p. 222).
18. Ullmer, B., Ishii, H., and Jacob, R.J. (2005). Token+constraint systems for tangible interaction with digital information. ACM Trans. Comput.-Hum. Interact., 12, 81-118.
19. Department for Digital, Culture, Media & Sport. *Secure by Design report* [online]. 2018. Available from https://www.gov.uk/government/publications/secure-by-design-report [Accessed 26 Sep 2020].
20. Rossi A., Haapio H. 'Proactive Legal Design: Embedding Values in the Design of Legal Artefacts' in In Schweighofer E., Kummer F., Saarenpää A. (eds.): *Internet of Things. Proceedings of the 22nd International Legal Informatics Symposium IRIS 2019*. Bern: Editions Weblaw; 2019. pp. 537–544.
21. Chen J., Edwards L., Urquhart L., McAuley D. 'Who is responsible for data processing in smart homes? Reconsidering joint controllership and the household exemption'. *International Data Privacy Law*. 2020.
22. Piasecki S., Urquhart L., McAuley D. *Defence Against Dark Artefacts: An Analysis of the Assumptions Underpinning Smart Home Cybersecurity Standards* 2019. http://dx.doi.org/10.2139/ssrn.3463799.
23. Lawrence E. Cohen and Marcus Felson, 'Social Change and Crime Rate Trends: A Routine Activity Approach' (1979) 44 American Sociological Review 588
24. Moral-IT Cards; Privacy by Design Cards.
25. Department for Digital, Culture, Media & Sport. *Code of Practice for Consumer IoT Security* [online]. 2018. Available from https://www.gov.uk/government/publications/code-of-practice-for-consumer-iot-security [Accessed 26 Sep 2020].
26. Department for Digital, Culture, Media & Sport. *Code of Practice for Consumer IoT Security* [online]. 2020. Available from https://www.gov.uk/government/consultations/consultation-on-regulatory-proposals-on-consumer-iot-security [Accessed 26 Sep 2020].
27. Department for Digital, Culture, Media & Sport. *Proposals for regulating consumer smart product cyber security - call for views* [online]. 2020. Available from https://www.gov.uk/government/publications/proposals-for-

regulating-consumer-smart-product-cyber-security-call-for-views [Accessed 26 Sep 2020].

28. Department for Digital, Culture, Media & Sport. *Cyber Security Incentives & Regulation Review: Call for Evidence* [online]. 2019. Available from https://www.gov.uk/government/publications/cyber-security-incentives-regulation-review-call-for-evidence [Accessed 26 Sep 2020].

29. McAuley D., Koene, A., Chen J. *Response to Consultation on the Government's Regulatory Proposals Regarding Consumer Internet of Things (IoT) Security* [online]. 2019. Available from https://doi.org/10.17639/4esm-9705

30. McAuley D., Haddadi, H., Urquhart L., Chen J. *Response to the Government's Call for Views: Proposals for Regulating Consumer Smart Product Cyber Security* https://nottingham-repository.worktribe.com/output/4880636

31. McAuley D., Chen J. *Response to DCMS Call for Evidence: Cyber Security Incentives and Regulation* https://doi.org/10.17639/SWKM-5T76.

32. Department for Digital, Culture, Media & Sport. *Consultation on the Government's regulatory proposals regarding consumer Internet of Things (IoT) security* [online]. 2020. Available from https://www.gov.uk/government/consultations/consultation-on-regulatory-proposals-on-consumer-iot-security/consultation-on-the-governments-regulatory-proposals-regarding-consumer-internet-of-things-iot-security [Accessed 26 Sep 2020].

33. Department for Digital, Culture, Media & Sport. *Proposals for regulating consumer smart product cyber security - call for views* [online]. 2020. Available from https://www.gov.uk/government/publications/proposals-for-regulating-

34. McAuley D., Koene, A., Chen J. *Response to Consultation on the Government's Regulatory Proposals Regarding Consumer Internet of Things (IoT) Security* [online]. 2019. https://doi.org/10.17639/4esm-9705